\documentstyle[12pt,epsfig]{article}
\def\cite#1{#1}
\newcommand{\ct}[1]{[\cite{#1}]}
\def\thebibliography#1{\section*{References}\list
 {[\arabic{enumi}]}{\settowidth\labelwidth{[#1]}\leftmargin\labelwidth
 \advance\leftmargin\labelsep
 \usecounter{enumi}}
 \def\newblock{\hskip .11em plus .33em minus -.07em}
 \sloppy
 \sfcode`\.=1000\relax}

\begin{document}

\begin{center}
{\Large\bf Contributions of $e^+e^-\rightarrow P(S)\gamma$
processes to muon $g-2$}\footnote{Contribution presented at Pisa
SIGHAD03, Workshop on Hadronic Cross Section at Low Energy,
October 8-10, 2003, Italy}
\end{center}

\begin{center}
A.-Z.~Dubni\v{c}kov\'a$^a$, S.~Dubni\v{c}ka$^b$, A.~Liptaj$^b$,
 R.~Pek\'arik$^b$
\end{center}

\begin{center} {
$^{a}$ \it Comenius Univ., Dept. of Theor. Physics, Faculty of
Mathematics, Physics and Informatics, 842 48 Bratislava,
Slovak Republic }\\
{$^{b}$ \it Inst. of Physics, Slovak Acad. of Sci., D\'ubravsk\'a
cesta 9,845 11 Bratislava,
Slovak Republic}\\
\end{center}

\begin{abstract}
The present situation on a comparison of theoretical evaluation of
the muon anomalous magnetic moment with experimental one is
roughly reviewed. Then by means of a recently elaborated unitary
and analytic model of the meson transition form factors the
contributions of $e^+e^-\to P(S) \gamma$ processes to muon $g-2$
is estimated.
\end{abstract}

\section{INTRODUCTION}

The muon magnetic moment is related to the spin by the expression
\begin{equation}
  \vec{\mu}=g\left({e\over 2m_\mu}\right)\vec{s} \label{z1}
\end{equation}
where the  gyromagnetic ratio $g$ is predicted theoretically (see
e.g. ref. \ct{1}) to be exactly  2.
   However, interactions existing in nature modify $g$
 to be slightly  exceeding value 2 because of the  emission and absorption of
 virtual photons, intermediate vector and Higgs bosons and vacuum polarization
 into virtual hadronic states.

In order to describe this modification of $g$ theoretically, the
magnetic anomaly was introduced by the relation
\begin{equation}
 a_\mu\equiv \frac{g-2}{2}=a_\mu^{(1)}\left({\alpha\over\pi}
  \right ) + \left(a_\mu^{(2)QED}\right.+  \left.a_\mu^{(2)had}\right )
  \left({\alpha\over\pi}\right)^2 +
   a_\mu^{(2)weak} + O\left({\alpha\over\pi}\right)^3 \label{z2}
\end{equation}
where $\alpha$ is the fine structure constant.

   The muon anomalous magnetic moment $a_{\mu}$ is
very interesting object for  theoretical and experimental
investigations. It is the best measured quantity in physics
\ct{2},\ct{3}
\begin{eqnarray}
  a_{\mu^+}^{exp}&=&(116 592 040 \pm 86)\times 10^{-11}\label{z3}
  \\
  a_{\mu^-}^{exp}&=&(116 592 140 \pm 85)\times 10^{-11}. \nonumber
\end{eqnarray}
On the other hand, an accurate theoretical (better
phenomenological) evaluation of $a_{\mu}^{th}$ provides
 an extremely  clean test of "Electroweak theory" and any
  disagreement between $a_{\mu}^{th}$ and $a_{\mu}^{exp}$ may give
  the first window to new physics beyond the Standard Model (SM).
  So, just the latter pretends the evaluation of $a_{\mu}^{th}$ to
  be as precise as possible, in order to be sure that SM is taken
  into account with sufficiently high precision.

  Nowadays the weak interaction contributions, arising from single
  and two-loop diagrams, are \ct{4}-\ct{6}
\begin{equation}
  a_\mu^{(2,3)weak}=(152 \pm 4)\times 10^{-11}.\label{z4}
\end{equation}
The  QED contributions up to 8-th order give the value \ct{7}
\begin{equation}
   a_{\mu}^{QED}=(116 584 705.7 \pm 2.9)\times10^{-11}.\label{z5}
\end{equation}

Though T. Kinoshita recently revealed in his calculations a
program error \ct{8}, the correction is not large enough to affect
the comparison between theory and experiment for the muon $g-2$,
but it only does alter the inferred value for the fine structure
constant.

The strong interactions contribute  through the  Feynman
 diagrams presented in Figs.1-2.
\begin{figure}[th]
\centering
\includegraphics[scale=0.4]{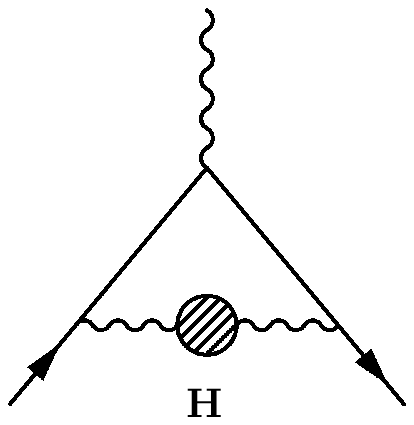}
\caption{The lowest-order hadronic vacuum-polarization
contributions.} \label{fig:1}
\end{figure}

\begin{figure}[th]
\centering
\includegraphics[scale=0.4]{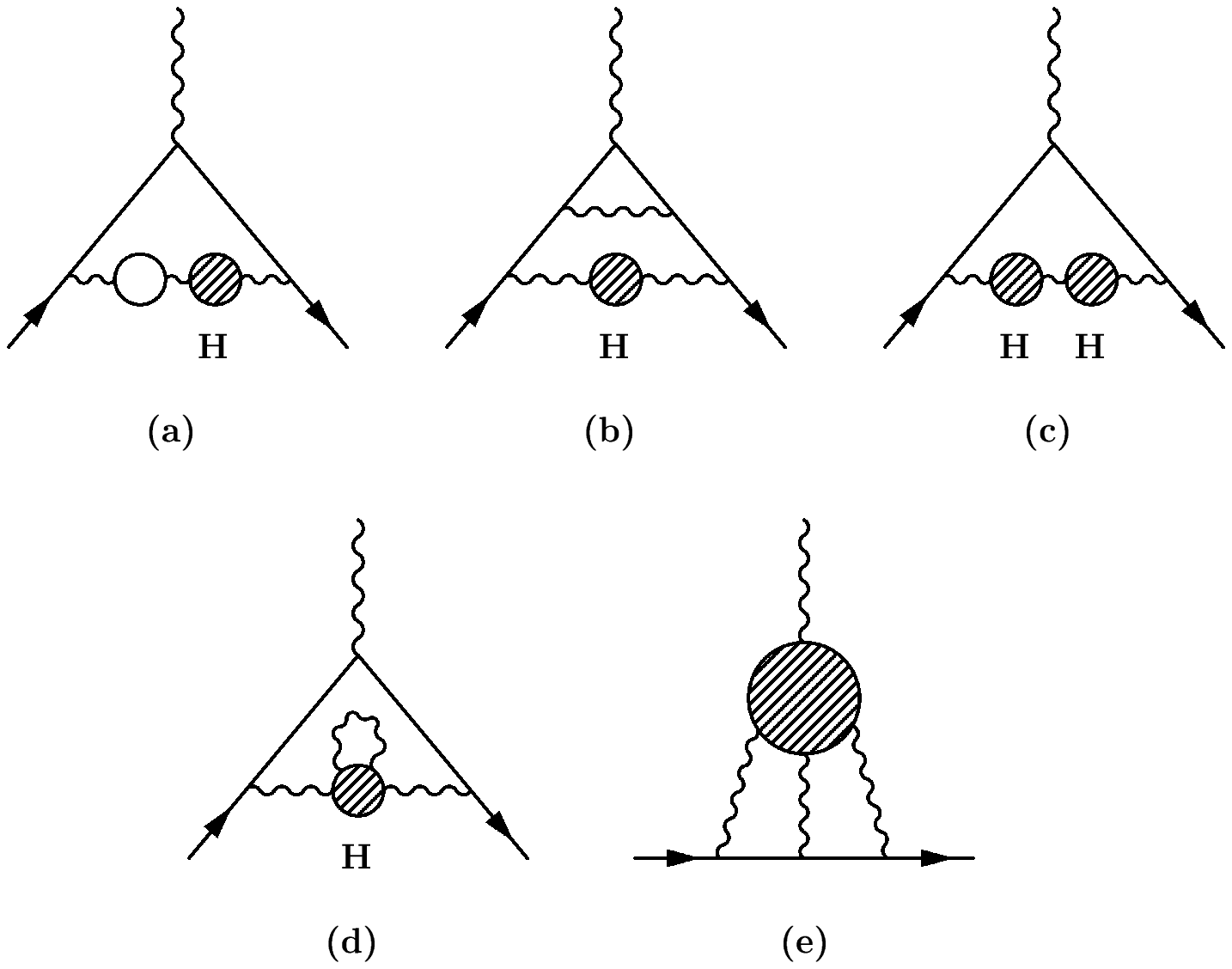}
\caption{The third-order hadronic vacuum-polarization
contributions.} \label{fig:2}
\end{figure}

The  lowest-order hadronic vacuum-polarization
   contributions (Fig.1), according to the more recent
    evaluations (considering only $e^+e^-$ annihilation cross-sections as there
    is still unsolved problem of disagreement \ct{9} with $\tau$-decay data)
\begin{eqnarray}
 a^{had}_{\mu}(l.o.) &=& (6 969.5\pm 80.7)\times 10^{-11}, \quad {\rm \ct{10}}
 \label{z6} \\
 a^{had}_{\mu}(l.o.) &=& (6 930.0\pm 71.7)\times 10^{-11}, \quad {\rm
 \ct{11}}\nonumber \\
  a^{had}_{\mu}(l.o.) &=& (6 961.5\pm 63.7)\times 10^{-11}, \quad {\rm
  \ct{12}}\nonumber
\end{eqnarray}
which seem to be more or less mutually consistent.

Within the recent past the  most discussed from all hadronic
contributions were the light--by--light (LBL) meson pole terms
(see Figs.2e and 3).
\begin{figure}[th]
\begin{center}
\includegraphics[scale=0.6]{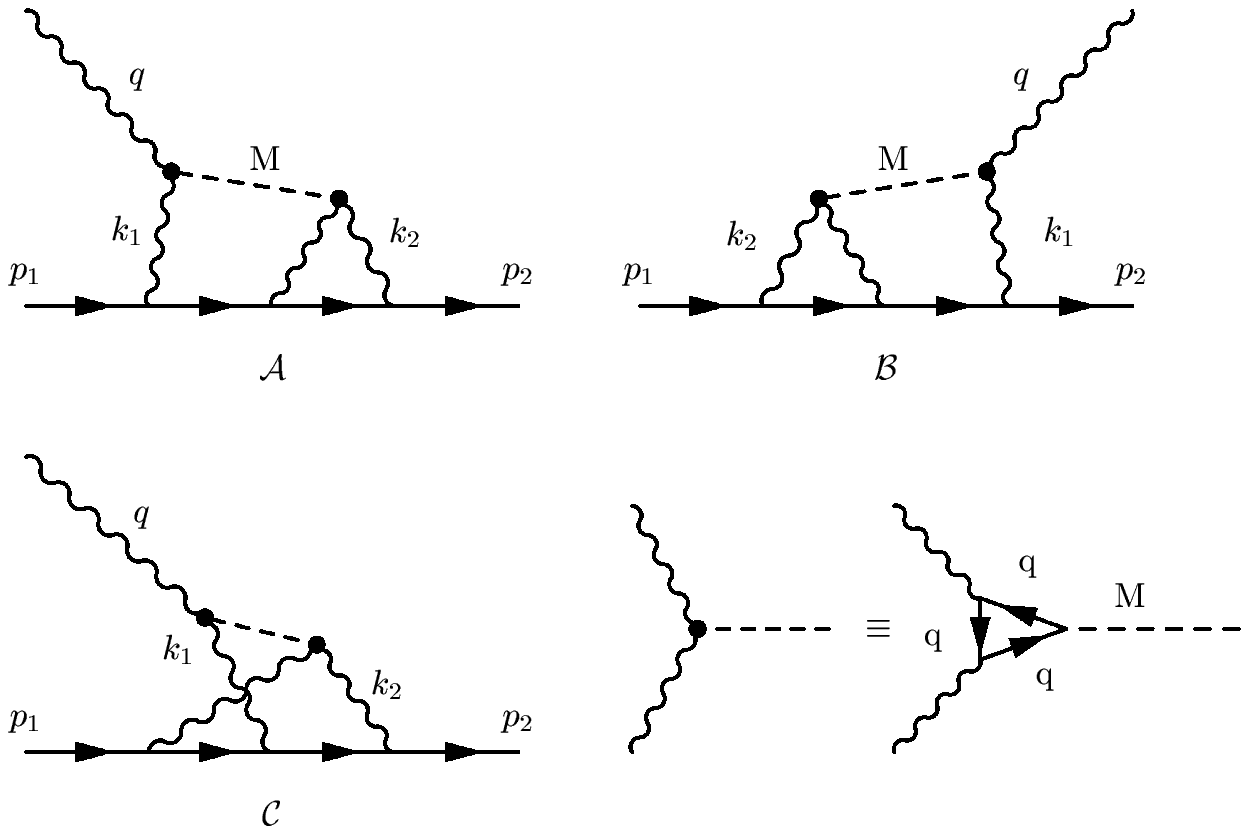}
\caption{Meson (M) pole diagrams in the third order hadronic
light--by--light scattering contributions to $a_\mu^{had}$.}
\label{fig:3}
\end{center}
\end{figure}

More precisely, up to the end of 2001 there was a believe  that
the light-by-light meson-pole term contributions are acquiring
negative values. However, recently it was clearly demonstrated
\ct{13}, \ct{14} that they must be positive ones, thus giving the
total LBL values
\begin{eqnarray}
a^{LBL}_{\mu} &=& (80\pm 40)\times 10^{-11} \quad {\rm \ct{15}}
\label{z7} \\
a^{LBL}_{\mu} &=& (111.2\pm 21.6)\times 10^{-11} \quad {\rm
\ct{16}} \nonumber \\
a^{LBL}_{\mu} &=& (136\pm 25)\times 10^{-11} \quad
{\rm\ct{17}}.\nonumber
\end{eqnarray}

\begin{figure}[ht]
\centering \includegraphics[scale=.4]{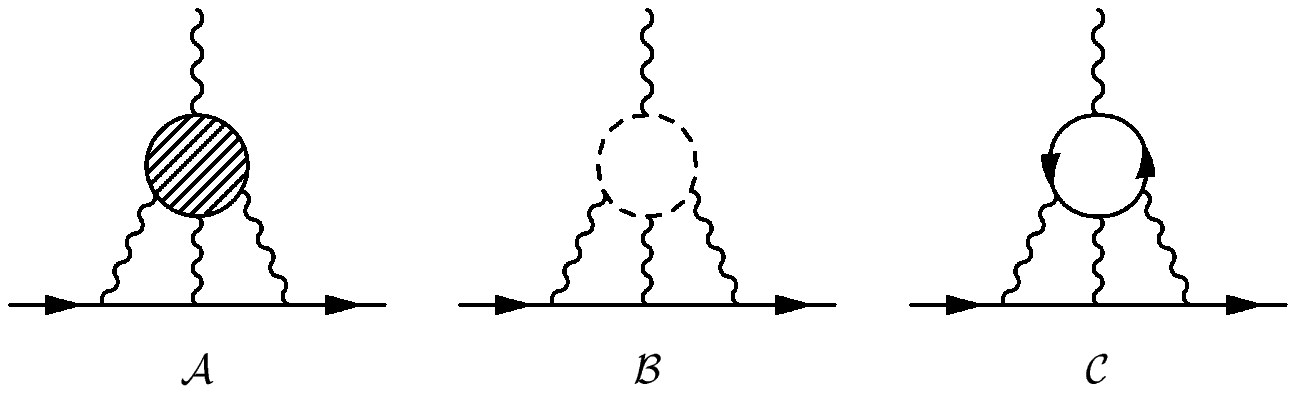} \caption{Third
order hadronic light--by--light scattering contribution to
$a_\mu^{had}$ ($\mathcal{A}$) and class of pseudoscalar meson
square loop diagrams ($\mathcal{B}$) and quark square loop
diagrams ($\mathcal{C}$) contributing to ($\mathcal{A}$).}
\label{fig:4}
\end{figure}

   The  others 3-loop hadronic contributions derived from the hadronic
vacuum polarizations $(VP)$ in Figs.2(a)-(c) aquire the value
\ct{18}
\begin{equation}
              a_\mu^{(3)VP} = (-101 \pm 6) \times
              10^{-11}.\label{z8}
\end{equation}

If we take into account $a_{\mu}^{LBL}$ from \ct{16}, which is
almost the average of all three ones in (\ref{z7}), then the total
3-loop hadronic correction is
\begin{equation}
       a_\mu^{(3)had} = (10.2 \pm 22.4) \times 10^{-11},\label{z9}
\end{equation}
where the errors have been added in quadratures.

Now,  summing all  presented  here contributions (\ref{z4}),
(\ref{z5}), the averaged (\ref{z6}) and (\ref{z9}), finally one
gets the SM theoretical prediction of the muon anomalous magnetic
moment value to be
\begin{equation}
              a_\mu^{th} = (116 591 832.6 \pm 82.7)
              \times 10^{-11},\label{z10}
\end{equation}
which in confrontation with the averaged experimentally determined
value (\ref{z3}) gives
\begin{equation}
              a_\mu^{exp} - a_\mu^{th} = (257.4 \pm 146.6) \times
              10^{-11}.\label{z11}
\end{equation}

This result might imply a window to the new physics beyond  SM.
   Therefore it is reasonable to pay more attention to exotic
   channels which could  diminish the difference in
 (\ref{z11}). Here we evaluate contributions of
 the   $e^+e^- \rightarrow P(S) \gamma$ processes (with $P$=$\pi^0$,
$\eta$, $\eta'$ and $S$=$ \sigma$, $a_0$), to muon $g-2$ by means
of the elaborated unitary and analytic model of transition form
factors (FF's) $F_{P(S)\gamma}(t)$ providing one analytic function
for space-like and time-like region simultaneously.

\section{$e^+e^- \rightarrow P\gamma$ PROCESSES}

There is a single FF $F_{P \gamma}(t)$ for each $\gamma^*\to
P\gamma$ transition completely describing  a behaviour of
\begin{equation}
\sigma_{tot}(e^+e^- \rightarrow P\gamma) = {{\pi{\alpha}^2}/2}(1 -
{m_P^2}/t)^3 {\mid F_{P\gamma}(t)\mid}^2 \label{z12}
\end{equation}
 to be defined by a parametrization of the matrix element of the
 electromagnetic (EM) current
 $J_{\mu}^{EM}=2/3\bar u\gamma_{\mu} u-1/3 \bar d \gamma_{\mu}
d-1/3 \bar s \gamma_{\mu} s$
\begin{equation}
\langle P(p)\gamma(k)\mid J_{\mu}^{EM}\mid 0\rangle =
\epsilon_{\mu\nu\alpha\beta} p^{\nu}\epsilon^{\alpha}k^{\beta}F_{
P \gamma}(t)\label{z13}\end{equation} where $\epsilon^{\alpha}$ is
the polarization vector of $\gamma$,
$\epsilon_{\mu\nu\alpha\beta}$ appears as only the pseudoscalar
meson belongs to the abnormal spin-parity series and
$t=q^2$=$-Q^2$ is the square four-momentum transferred by the
virtual photon. In the framework of the unitary and analytic model
of the EM pseudoscalar-meson-transition FF's \ct{19}
$F_{P\gamma}(t)$ takes form
\begin{equation}
F_{P \gamma }(t)=F_{P\gamma }^{I=0}[V(t)]+F_{P\gamma }^{I=1}[W(t)]
\label{z14}\end{equation} with
$$
F_{P \gamma
}^{I=0}[V(t)]=\left(\frac{1-V^2}{1-V_N^2}\right)^2\left
\{\frac{1}{2}F_{P \gamma}(0)H(\omega')\right.+
$$
$$
+\left.[L(\omega)-H(\omega')]a_{\omega}^P+ \right.
$$
$$
+[L(\phi)-H(\omega')]a_{\phi}^P\biggl \}
$$
$$
F_{P\gamma}^{I=1}[W(t)]=\left(\frac{1-W^2}{1-W_N^2}\right)^2\left
\{ \frac{1}{2}F_{P \gamma}(0)H(\rho')\right.+
[L(\rho)-H(\rho')]a_{\rho}^P\biggl \},
$$
where $V(t)$($W(t)$) is a  conformal mapping
\begin{equation} V(t)=i\frac
{\sqrt{q_{in}^{I=0}+q}-
 \sqrt{q_{in}^{I=0}-q}}
{\sqrt{q_{in}^{I=0}+q} + \sqrt{q_{in}^{I=0}-q}} \label{z15}
\end{equation}
$$
q=[(t-t_0)/t_0]^{1/2}; \quad
q_{in}^{I=0}=[(t_{in}^{I=0}-t_0)/t_0]^{1/2}$$ of the four-sheeted
Riemann surface in $t$-variable into one  $V$-plane ($W$-plane),
\begin{equation}
F_{P\gamma}(0)=\frac{2}{\alpha m_P}\sqrt{\frac{\Gamma(P\to \gamma
\gamma)}{\pi m_P}}\label{z16}\end{equation}
 $t_0=m_{\pi^0}^2$, $t_{in}^{I=0}$ and $t_{in}^{I=1}$ are
 the effective square-root branch points  including in average
contributions of all higher important thresholds in both,
isoscalar and isovector case, respectively,
$$
L(s)=
\frac{(V_N-V_{s})(V_N-V_{s}^*)(V_N-1/V_{s})(V_N-1/V_{s}^*)}{(V-V_{s})(V-V_{s}^*)(V-1/V_{s})(V-1/V_{s}^*)};
$$
$$
s=\omega,\phi, \quad V_N=V(t)_{|_{t=0}}
$$
$$
H(\omega')=\frac{(V_N-V_{\omega'})(V_N-V_{\omega'}^*)(V_N+V_{\omega'})(V_N+V_{\omega'}^*)}{(V-V_{\omega'})(V-V_{\omega'}^*)(V+V_{\omega'})(V+V_{\omega'}^*)};
$$
$$
L(\rho)=\frac{(W_N-W_{\rho})(W_N-W_{\rho}^*)(W_N-1/W_{\rho})(W_N-1/W_{\rho}^*)}{(W-W_{\rho})(W-W_{\rho}^*)(W
-1/W_{\rho})(W-1/W_{\rho}^*)};$$
$$
  W_N=W(t)_{|_{t=0}}
$$
$$
H(\rho')=\frac{(W_N-W_{\rho'})(W_N-W_{\rho'}^*)(W_N+W_{\rho'})(W_N+W_{\rho'}^*)}{(W-W_{\rho'})(W-W_{\rho'}^*)(W
+W_{\rho'})(W+W_{\rho'}^*)} $$  and
$$
 a_r^P=(f_{r P \gamma}/f_r);
\quad r=\rho, \omega, \phi.
$$

Prediction of  $F_{\pi^0\gamma}(t)$, $F_{\eta\gamma}(t)$
$F_{\eta'\gamma}(t)$ behaviours and their comparison with existing
data is presented in Figs. 5-7. By substitution of the latter into
(\ref{z12}) one estimates, through the first integral in the
relation
\begin{eqnarray}
a_{\mu}^{(2)had}&=&\frac{1}{4\pi^3}\left
\{\int_{\pi^0}^{3GeV^2}\sum_F\sigma_{tot}(e^+e^-\to
F)\right.\times K_{\mu}(t)dt+ \label{z17} \\
&+&\left.\int_{3GeV^2}^{\infty}R(e^+e^-\to
had)\sigma_{tot}(e^+e^-\to \mu^+\mu^-)\right.\times
K_{\mu}(t)dt\biggl \}\nonumber
\end{eqnarray}
with
\begin{equation}
K_{\mu}(t)=\int_0^1\frac{x^2(1-x)}{x^2+(1-x)t/m_\mu^2}dx,
\label{z18}
\end{equation}
which represents the diagram in Fig. 1, the contributions of
$e^+e^-\to P\gamma$ processes
\begin{eqnarray}
a_{\mu}(\pi^0\gamma)&=&17.2 \times 10^{-11}\nonumber \\
a_{\mu}(\eta\gamma)&=&2.2 \times 10^{-11} \label{z19} \\
a_{\mu}(\eta'\gamma)&=&1.5 \times 10^{-11}\nonumber \end{eqnarray}
into the muon anomalous magnetic moment $a_\mu$. The obtained
results don't agree with the values recently evaluated in papers
\ct{20} and \ct{21} and have to be carefully reanalyzed.

\section{ $e^+e^- \to S \gamma$ PROCESSES}

Similarly to the unitary and analytic model of $F_{P \gamma }(t)$
FF's, one can construct the  model for $F_{S\gamma }(t)$ FF's
\begin{equation}
F_{S\gamma}(t)=F_{S \gamma}^{I=0}[V(t)]+F_{S
\gamma}^{I=1}[W(t)]\label{z20}
\end{equation}
$$
F_{S \gamma}^{I=0}[V(t)]=\left(\frac{1-V^2}{1-V_N^2}\right)^2\left
\{\frac{1}{2}F_{S \gamma}(0)H(\omega')+\right.$$
$$+[L(\omega)-H(\omega')]a_{\omega}^S
+[L(\phi)-H(\omega')]a_{\phi}^S\biggl \}
$$
$$
F_{S \gamma}^{I=1}[W(t)]=\left(\frac{1-W^2}{1-W_N^2}\right)^2\left
\{ \frac{1}{2}F_{S \gamma}(0)H(\rho')\right.+
$$
$$
+ [L(\rho)-H(\rho')]a_{\rho}^S\biggl \}
$$
\begin{table*}[htb]
\caption{Vector meson-Scalar meson-Photon coupling constants.}
\label{table:1}
\newcommand{\m}{\hphantom{$-$}}
\newcommand{\cc}[1]{\multicolumn{1}{c}{#1}}
\renewcommand{\tabcolsep}{0.5pc} 
\renewcommand{\arraystretch}{1.0} 
\begin{tabular}{llllll} \hline
 g [GeV$^{-1}$]
     & {QCD sum rules (s.r.)} & Light cone QCD s.r & $\rho$ photoprod. &  chir.loops &{$\bar g$}
     \\ \hline
 $\rho\sigma\gamma$  & 4.15$\pm$ 0.78\ct{24} & 2.85$\pm$
0.52\ct{27} & 3.51\ct{30} &-& 3.5 \\  $\rho a_0
\gamma$ & 1.69$\pm$ 0.39 \ct{25}& - & - & -&1.69 \\
$\omega\sigma\gamma$ &
-& 0.92$\pm$ 0.10\ \ct{28}&-  & $0.14\pm 0.01$\ct{32}  &0.53 \\
 $\omega a_0\gamma$        & 0.57$\pm$ 0.13\ct{25} & - & - & - &
0.57 \\  $\phi\sigma\gamma$    & $0.042\pm
0.009$\ct{26}& 0.039$\pm$ 0.009\ct{29} & 0.046\ct{31} & - &0.042 \\
 $\phi a_0\gamma$    & 0.12$\pm$ 0.03\ct{26} & 0.13$\pm$
0.03\ct{29} & - & - &0.125\\ \hline
\end{tabular}\\[2
pt]
\end{table*}
\begin{figure}[htp] 
\centering
\includegraphics[scale=.4
]{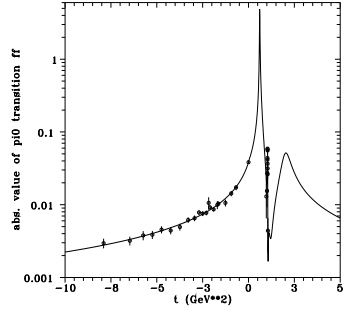} \caption{$\pi^0$ transition form factor.}\label{fig:5}
\end{figure}

\begin{figure}[htp] 
\centering
\includegraphics[scale=.4
]{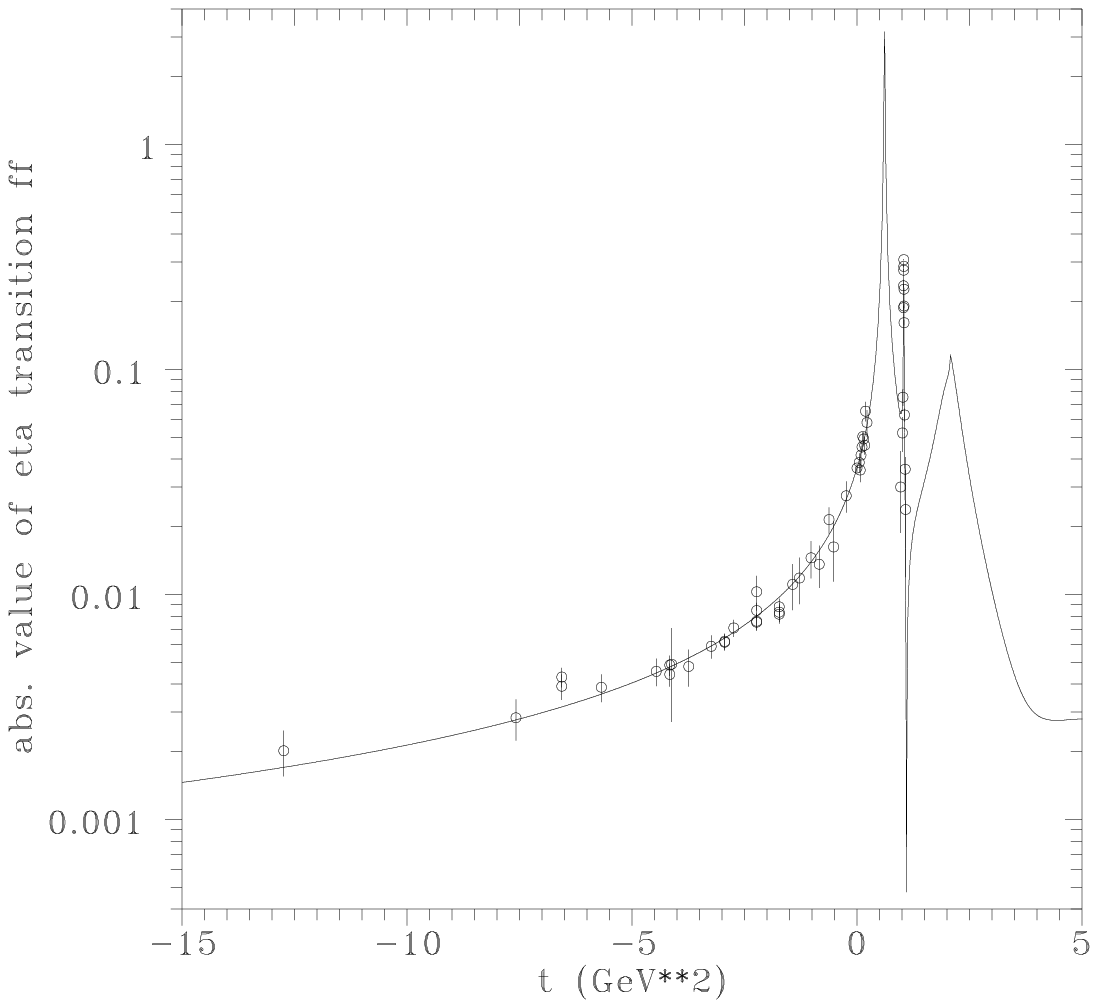} \caption{$\eta$ transition form factor.}\label{fig:6}
\end{figure}

\begin{figure}[htp] 
\centering
\includegraphics[scale=.5
]{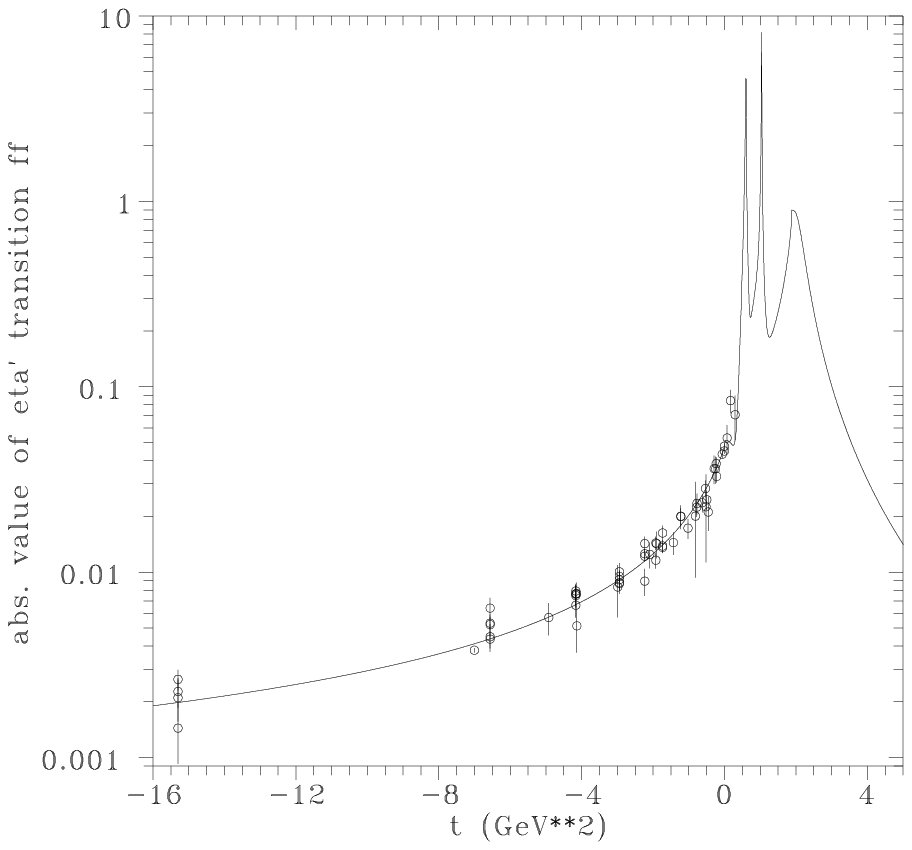} \caption{$\eta'$ transition form
factor.}\label{fig:7}
\end{figure}
with the same denotation as in the case of the pseudoscalar
mesons, however, here $t_0^{I=0}=9m_\pi^2$ and
$t_0^{I=1}=4m_\pi^2$. The effective branch points $t_{in}^{I=0}$
and $t_{in}^{I=1}$ are fixed at the typical value 1GeV$^2$ found
in a fitting of the pseudoscalar-meson transition FF data, which
corresponds approximately to the $K\bar K$ threshold. $F_{\sigma
\gamma}(0)$ value is found from the two-photon decay rate
$\Gamma(\sigma\to\gamma\gamma)$=0.283 keV estimated by some of us
\ct{22} in the framework of the Nambu-Jona-Lasinio model
(\ref{z7}), taking into account the fact that $g_{\sigma q\bar
q}\equiv g_{\pi^0 q\bar q}$. $F_{a_0\gamma}(0)$ is determined from
the experimental value $\Gamma(a_0\to \gamma\gamma)$=0.24 keV
given in \ct{23}. The coupling constant ratios
$a_r^S$=$(f_{rS\gamma}/f_r)$, $r$=$\sigma$, $a_0$ are determined
by using the averaged values $\bar{g}\equiv f_{rS\gamma}$
presented in Table 1. Then the behaviours of $|F_{S\gamma}(t)|$
are found (see dashed lines Figs. 8,9) and subsequently the
contributions of $e^+e^-\to S \gamma$ processes to $a_{\mu}^{th}$
are evaluated $a_{\mu}(\sigma\gamma)=1081.8\times 10^{-11}$,
$a_{\mu}(a_0\gamma)=84.5\times 10^{-11}$, by means of the first
integral in (\ref{z17}).The latter clearly indicate that values of
coupling constants in Table 1 are overestimated. This our
hypothesis is confirmed in the case of $f_{\rho\sigma\gamma}$ by
Rekalo and Tomasi-Gustafsson \ct{33}, who found the latter to be
10-times smaller than in Table 1.
\begin{figure}[htp] 
\centering
\includegraphics[scale=.5
]{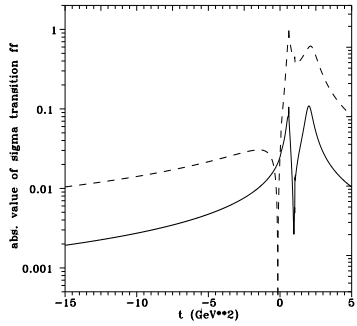} \caption{$\sigma$ transition form
factor.}\label{fig:8}
\end{figure}
\begin{figure}[htp] 
\centering
\includegraphics[scale=.5
]{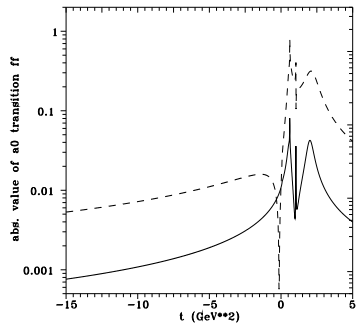} \caption{$a_0$ transition form
factor.}\label{fig:9}
\end{figure}
If we take in the lump all averaged values $\bar g$ of
vector-meson scalar-meson photon coupling constants from Table 1
10-times smaller, one obtaines the behaviours of
$|F_{\sigma\gamma}(t)|$ and $|F_{a_0\gamma}(t)|$  with smooth fall
in the space-like region (see full lines in Figs. 8,9) and quite
reasonable values
\begin{eqnarray}
a_{\mu}(\sigma\gamma)&=&12.46 \times 10^{-11} \label{z21}\\
a_{\mu}(a_0\gamma)&=&0.92 \times 10^{-11} \nonumber
\end{eqnarray}
of $e^+e^-\to S\gamma$ contributions to $a_{\mu}^{th}$.

\section{CONCLUSIONS}

By exploiting recently elaborated unitary and analytic model of
P-meson transition FF's, we have evaluated realistic contributions
of $e^+e^-\to P \gamma$ ($P=\pi^0$, $\eta$, $\eta'$) processes to
$a_{\mu}^{th}$.

The same model was applied also to the evaluation of contributions
of $e^+e^-\to S\gamma$ ($S$=$\sigma$, $a_0$) processes to
$a_\mu^{th}$. However, in this case according to our knowledge
there is no experimental information on corresponding transition
FF's up to now and the predicted $e^+e^-\to S \gamma$
contributions (\ref{z21}) to $a_\mu^{th}$ are more or less
hypothetical which should be somehow justified.

This work was in part supported by Slovak Grant Agency for
Sciences, Grant 2/1111/04 (A.Z.D, S.D., A.L, R.P.).


\begin{thebibliography}{9}
\bibitem{Ait03} I.J.R. Aitchison and A.J.G. Hey: Gauge theories in
particle physics, Institute of Physics Publishing, Briston and
Philadelphia, 2003.
\bibitem{Ben02} G.W. Bennet et al, Phys. Rev. Lett. 89 (2002)
101804-1.
\bibitem{Bennet}
G.W. Bennet et al, hep-exp/0401008.
\bibitem{Degr98} G. Degrassi, G.F. Giudice, Phys. Rev. D58 (1998)
53007.
\bibitem{Czar99} A. Czarnecki, W. Marciano, Nucl. Phys. B (Proc.
Suppl.) 76 (1999) 245.
\bibitem{Knecht02} M. Knecht, S. Peris, M. Perrottet, E. de
Rafael, JHEP 0211 (2002) 003.
\bibitem{Hughes99} V.H. Hughes, T. Kinoshita, Rev. Mod. Phys. 71
(1999) S133.
\bibitem{Kino03} T. Kinoshita, M. Nio, Phys. Rev. Lett. 90 (2003)
021803.
\bibitem{michel}
M. Davier, S. Eidelman, A. H\"ocker, Z. Zhang, Eur.Phys. J. C27
(2003) 497.
\bibitem{Naris01} S. Narison, Phys. Lett. 513B (2001) 53.
\bibitem{Miche}
M. Davier, S. Eidelman, A. H\"ocker, Z. Zhang, Eur.Phys. J. C31
(2003) 503.
\bibitem{Hgiwara}
K. Hgiwara, A.D. Martin, D. Nomura, T. Teubner, hep-ph/0312250.
\bibitem{knecht}
M. Knecht, A. Nyffeler, Phys. Rev. D65 (2002) 073034.
\bibitem{Kne02} M. Knecht, A. Nyffeler, M. Perrottet, E. de
Rafael, Phys. Rev. Lett. 88 (2002) 071802.
\bibitem{Nyfeler}
A. Nyffeler, hep-ph/0203347.
\bibitem{Bart02} E. Barto\v s, A.Z. Dubni\v ckov\'a, S. Dubni\v
cka, E.A. Kuraev, E. Zemlyanaya, Nucl. Phys. B632 (2002) 330.
\bibitem{Melnikov}
K. Melnikov, hep-ph/0312226.
\bibitem{Krause97} B. Krause, Phys. Lett. B390 (1997) 392.
\bibitem{Dubn03} A.Z. Dubni\v ckov\'a, S. Dubni\v cka, G.
Pancheri, R. Pek\'arik, Contr. at Int. Conf. PHOTON'03, April
7-11, 2003, Frascati, (Roma) Italy, Nucl. Phys. B (Proc,Suppl.)
126 (2004) 71.
\bibitem{achasov}
N.N. Achasov, A.V. Kiselev, Phys. Rev. D65 (2002) 097302.
\bibitem{troconiz}
J.F. de Troconiz, F.J. Yndurain, Phys. Rev. D65 (2002) 093001.
\bibitem{Dubni} S. Dubni\v cka, A.Z. Dubni\v ckov\'a, M. Se\v
cansk\'y (to be published).
\bibitem{Hagi02} K. Hagiwara, et al, Phys. Rev. D66 (2002) 010001.
\bibitem{Goka1}A. Gokalp, O. Yilmaz, Phys. Rev. D64 (2001) 034012.
\bibitem{Goka2}A. Gokalp, O. Yilmaz, Eur. Phys. J. C22 (2001) 323.
\bibitem{Goka3}A. Gokalp, O. Yilmaz, Acta Phys. Polonica B33 (2002) 1313.
\bibitem{Aliev02}T.M. Aliev et al, Phys. Rev. D65 (2002) 076004.
\bibitem{Goka4}A. Gokalp, O. Yilmaz, hep-ph/0202091 (2002).
\bibitem{Goka5}A. Gokalp, O. Yilmaz, hep-ph/0111072 (2001).
\bibitem{Fri96}B. Friman, M. Soyeur, Nucl. Phys. A600 (1996) 477.
\bibitem{Titov99}A.I. Titov et al, Phys. Rev. C60 (1999) 035205.
\bibitem{Goka6}A. Gokalp et al, Phys. Rev. D67 (2003) 073008.
\bibitem{Reka03} M. Rekalo, E. Tomasi-Gustafsson, Nucl. Phys. A714
(2003) 632.
\end{thebibliography}
\end{document}